# Incorporating energy storage and user experience in isolated microgrid dispatch using a multi-objective model

Yang Li [1,2,*], Zhen Yang [1], Dongbo Zhao [2], Hangtian Lei [3], Bai Cui [2], Shaoyan Li [4]

[1] School of Electrical Engineering, Northeast Electric Power University, Jilin 132012, China
[2] Energy Systems Division, Argonne National Laboratory, Lemont, IL 60439, USA
[3] Department of Electrical and Computer Engineering, University of Idaho, Moscow, ID 83844, USA
[4] School of Electrical and Electronic Engineering, North China Electric Power University, Baoding 071003, China

[*] Corresponding author (Yang Li). Email: liyang@neepu.edu.cn

**Abstract:** In order to coordinate multiple different scheduling objectives from the perspectives of economy, environment and users, a practical multi-objective dynamic optimal dispatch model incorporating energy storage and user experience is proposed for isolated microgrids. In this model, besides Microturbine units, energy storage is employed to provide spinning reserve services for microgirds; and furthermore, from the perspective of demand side management, a consumer satisfaction indicator is developed to measure the quality of user experience. A two-step solution methodology incorporating multi-objective optimization (MOO) and decision analysis is put forward to address this model. First, a powerful heuristic optimization algorithm, called the $\theta$-dominance based evolutionary algorithm, is used to find a well-distributed set of Pareto-optimal solutions of the problem. And thereby, the best compromise solutions (BCSs) are identified from the entire solutions with the use of decision analysis by integrating fuzzy C-means clustering and grey relation projection. The simulation results on the modified Oak Ridge National Laboratory Distributed Energy Control and Communication lab microgrid test system demonstrate the effectiveness of the proposed approach.

## NOMENCLATURE

**Acronyms**

| | |
|---|---|
| MGs | Microgrids |
| IMG | Isolated microgrid |
| ESS | Energy storage system |
| DERs | Distributed energy resources |
| DGs | Distributed generations |
| MILP | Mixed integer linear programming |
| PDF | Probability density function |
| SOT | sequence operation theory |
| ORNL | Oak Ridge National Laboratory |
| DECC | Distributed energy control and communication |
| TOU | Time-of-use |
| DSM | Demand side management |
| MOO | Multi-objective optimization |
| SRs | Spinning reserves |
| BCSs | Best compromise solutions |
| NSGA-II | Non-dominated sorting genetic algorithm II |
| $\theta$-DEA | $\theta$-dominance based evolutionary algorithm |
| POSs | Pareto optimal solutions |
| FCM | Fuzzy C-means |
| GRP | Grey relation projection |
| PSs | Probabilistic sequences |
| RPV | Relative projection value |

**Symbols**

| | |
|---|---|
| $q$ | Discrete step size (kW) |
| $F_1$ | IMG operation cost ($) |
| $F_2$ | Gas emission (g/kWh) |
| $F_3$ | Consumer satisfaction (%) |
| $t$ | A scheduling time period (h) |
| $T$ | Total number of time periods in a cycle (h) |
| $\eta_{ch}$ | Charge coefficient (p.u) |
| $\eta_{dc}$ | Discharge coefficient (p.u) |
| $P_{Ress}$ | Reserve capacity of ESS |
| $\omega_{rt\_price}$ | TOU price |
| $\omega_{rc\_price}$ | SR price |
| MG | Total number of MT units |
| C | Capacity (kWh) |
| $\alpha$ | Pre-given confidence level (%) |

**Subscripts**

| | |
|---|---|
| $w$ | Wind |
| $*$ | Rated |
| $r$ | Actual light intensities |
| $max$ | Maximum value |
| $min$ | Minimum value |
| pv | Photovoltaic |
| $a$ | Probabilistic sequences of WT power outputs |
| $b$ | Probabilistic sequences of PV power outputs |
| $c$ | Probabilistic sequences of joint power outputs of PV and WT |
| $d$ | Load probabilistic sequences |
| $e$ | Equivalent load probabilistic sequences |
| $n$ | Number of MT |
| $rob$ | Probability |
| Ress | Reserve capacity |
| L | Load |
| $k$ | Different kinds of pollution gases |
| $\mu$ | Degree of membership |

**Superscripts**

| CH | Charge |
| DC | Discharge |
| WT | Wind turbine |
| PV | Photovoltaic |
| MT | Microturbine |
| EL | Equivalent load |
| PRO | Projection of a scheduling scheme |
| GRC | Grey relation coefficient |

## 1. Introduction

As a locally controlled system including interconnected loads and distributed generations (DGs), a microgrid (MG) is able to connect or disconnect from the traditional centralized grid, enabling it to operate flexibly and efficiently in both grid-connected or island-modes [1, 2]. Previous research has demonstrated that MGs are capable of improving the receptivity of distribution systems to distributed energy resources (DERs) and enhancing the efficiency of renewable energy utilization [3-5]. In [3], optimal power dispatch of DGs in MGs under uncertainties is formulated and solved by using the imperialist competitive algorithm. In [4], a cooperative game approach is presented to coordinate multi-microgrid operation within distribution systems. In [5], microgrids are employed to provide ancillary services of voltage control for distribution networks. Compared with traditional distribution networks, microgrids are of significant advantages in reliability, economy, and self-healing [6], and such systems are generally designed to provide power for small communities or industries [7]. In extreme weather-related incidents or inaccessible to the main power grid, an isolated MGs (IMG) plays a unique role in guaranteeing an uninterruptible power supply to critical loads [8, 9]. In addition, it can also be used to supply power to users in remote areas involving rural villages, islands, and deserts [10]. For example, an intelligent control technique is proposed for rural isolated village MGs by using multi-agent modelling and price-based demand response in [11]. However, there are still some open problems in finding the optimal operation schemes with reasonable fuel costs, gas emissions, and consumer satisfaction, while satisfying a series of variously related constraints. At the same time, demand side resources are of growing importance in the successful market integration of renewable energy sources [12], and price-based demand responses such as the real-time pricing [13] and the peak-valley time-of-use (TOU) pricing [14] have recently proven to be significant measures to implement demand side management (DSM) in MGs [15]. In [13], a MG scheduling model is proposed to coordinate IMG and electric vehicle battery swapping station in multi-stakeholder scenarios via real-time pricing. In [14], a TOU tariff algorithm is developed for residential prosumer DSM. In [15], DSM is applied to cover the uncertainty of renewable generations by using demand response.

A lot of researches on MGs have been undergoing a boom around the world for the last past years after demonstrating its great value in critical situations with a very wide scope covering various aspects such as planning, operation [16, 17], and control strategies [18]. In [16], a day-ahead optimal scheduling method is presented for grid-connected MGs by using energy storage control strategy. In [17], taking into account demand response and the uncertainty in the generation and load, a decentralized algorithm for energy trading among the load aggregators and generators has been proposed, and the extensive simulation results on different standard test feeders demonstrate the effectiveness and superiority of the approach. In [18], S-shaped droop control method with secondary frequency characteristics is developed for inverters in MGs. The optimal scheduling problem of isolated MG with multiple objectives is paid attention to in this research. The operation of a MG is vulnerable because of its small system capacity and the uncertainties from distributed generations and loads [1]. It is also difficult for operators to coordinate effectively multi-objective functions in the process of optimization, simultaneously. It is quite important for MG optimal scheduling to resolve these key problems.

In order to address these challenges, various means have been adopted in previous works, reference [19] presents an integrated dispatching method based on robust multi-objective to improve the economy and environmental benefits for a microgrid. A new multi-objective optimization (MOO) approach is presented in [20] to find the minimum value of the annualized cost expected load loss and energy loss based on a hybrid wind-solar generation microgrid system. Reference [21] puts forward a modified multi-objective intelligent evolutionary algorithm for coordinating the economic/environmental objectives for microgrids. A two-step power scheduling approach is presented to resolve the multi-objective optimal planning problem of isolated MGs using different benefit subjects in [22]. A comprehensive MG scheduling strategy is presented to reduce imbalances between economic and technological objectives in [23]. A multi-objective load scheduling model has been presented for microgrids incorporating DGs and electric vehicles in [24]. In [25], a multi-objective stochastic programming is utilized for the MG energy management by considering demand responses. To balance cost and reliability, the non-dominated sorting genetic algorithm II (NSGA-II) has been developed for solving the optimal sizing issue of a DC microgrid in [26]. Reference [27] proposes an intelligent energy management formulation of microgrids based on artificial intelligence techniques and employs MOO techniques to seek the minimum value of the operation cost and the environmental impact for microgrids. In [28], a new multi-objective optimal scheduling method with security constraints is presented based on Pareto Concavity Elimination Transformation, a MOO technique to minimize the generation cost and the reliability cost.

However, there are still some research gaps in this area. (1) In previous works, dispatchable generators such as microturbines (MTs) are usually used to supply spinning reserves (SRs) for a microgrid, but recent investigations show an energy storage system (ESS) may be a preferable option for doing so due to its faster responding and lower cost [8]. (2) At the same time, most of the existing work in this area gives much focus to take into account the economic and environmental benefits of MG dispatch, and relatively less attention to user experience, which makes the obtained MG dispatch scheme may not meet the growing diverse needs of users. (3) Determination of best compromise solutions (BCSs) from all the Pareto optimals is a tricky issue in practical applications, not only because decision-makers often have inconsistent choices on the generated numerable solutions, but also because, even for the same decision-maker, his decision preferences may change according to the actual MG operational requirements.

To address the above problems, this work incorporates energy storage and user experience in isolated microgrid dispatch using a multi-objective model. The contributions of this study mainly include the following aspects:

- Dispatch model: A multi-objective dynamic optimal dispatch model incorporating energy storage and user experience is proposed for isolated microgrids. In this model, besides Microturbine units in existing approaches, energy storage is employed to provide spinning reserve services for microgirds; and a consumer satisfaction indicator is developed to measure the quality of user experience.
- Two-step solution methodology: The first step employs a powerful heuristic optimization algorithm called $\theta$-dominance based evolutionary algorithm ($\theta$-DEA) to generate a well-distributed set of Pareto-optimal solutions (POSs); and then, the best compromise solutions (BCSs) are determined from the entire solutions with the use of decision analysis by integrating fuzzy C-means clustering (FCM) and grey relation projection (GRP).
- The simulation results on a concrete microgrid test system demonstrate that the proposed approach manages to solve the dispatch model and automatically yields the BCSs representing decision makers' different preferences. In addition, its computational efficiency meets the real-time requirements of microgrid scheduling.

This article is structured as follows. The uncertainty modeling of MG is introduced in Section 2; and then, Section 3 gives the problem formulation in detail; in the next place, its further solution approach is detailedly illustrated in the next section. Simulations tests are carried out in Section 5, and Section 6 draws the conclusions and points out the possible future directions.

## 2. The model of MG

### 2.1. WT Model

It's known that the wind speed obeys the Weibull distribution [29]. And thereby, previous research shows that the probability density function (PDF) of wind speed can be calculated by

$$f_w(v) = (k/\gamma)(v/\gamma)^{k-1}\exp[-(v/\gamma)^k] \quad (1)$$

where $v$, $k$ and $\gamma$ are respectively the real wind speed, the shape factor and the scale factor.

The PDF of the WT output $f_o(P^{WT})$ is [29]

$$f_o(P^{WT}) = \begin{cases} (khv_{in}/\gamma P_*)\left[((1+hP^{WT}/P_*)v_{in})/\gamma\right]^{k-1} \times \\ \exp\left\{-\left[((1+hP^{WT}/P_*)v_{in})/\gamma\right]^k\right\}, p^{WT} \in [0, P_*] \\ 0, \quad \text{otherwise} \end{cases} \quad (2)$$

where $h = (v_*/v_{in}) - 1$.

### 2.2. PV Model

According to previous research, the PDF of solar irradiance follows the following beta distribution:

$$f_r(r) = \frac{\Gamma(\lambda_1)+\Gamma(\lambda_2)}{\Gamma(\lambda_1)\Gamma(\lambda_2)}\left(\frac{r}{r_{max}}\right)^{\lambda_1-1}\left(1-\frac{r}{r_{max}}\right)^{\lambda_2-1} \quad (3)$$

where $r_{max}$ is the maximum value of the actual light intensity $r$; $\lambda_1$ and $\lambda_2$ are the shape factors; $\Gamma$ is a Gamma function. The PV outputs can be obtained according to the following formula [29]:

$$P^{PV} = \xi A^{PV}\eta^{PV} \quad (4)$$

where $\xi$ is the solar irradiance, $A^{PV}$ and $\eta^{PV}$ are respectively the radiation area and the conversion efficiency of this PV.

The PDF of PV output is [8, 29]

$$f_p(P^{PV}) = \frac{\Gamma(\lambda_1)+\Gamma(\lambda_2)}{\Gamma(\lambda_1)\Gamma(\lambda_2)}\left(\frac{P^{PV}}{P^{PV}_{max}}\right)^{\lambda_1-1}\left(1-\frac{P^{PV}}{P^{PV}_{max}}\right)^{\lambda_2-1} \quad (5)$$

where $P^{PV}_{max}$ is the maximum value of the PV output $P^{PV}$.

### 2.3. Load Model

Assuming that load fluctuations follow the normal distribution, the PDF of loads is [30]

$$f_l(P^L) = \frac{1}{\sqrt{2\pi}\sigma_L}\exp\left(-\frac{(P^L-\mu_L)^2}{2\sigma_L^2}\right) \quad (6)$$

where $P^L$ denotes the load power; while $\mu_L$ and $\sigma_L$ are respectively its mean value and the standard deviation.

### 2.4. Equivalent Load Model

For ease of analysis, an equivalent load (EL) power is defined by [8]

$$P^{EL} = P^L - (P^{WT} + P^{PV}) \quad (7)$$

where $P^{EL}$ is the predictive value of the EL power.

## 3. Optimal Scheduling Model

### 3.1. Objective Functions

#### 3.1.1 IMG Operation Cost:
In this study, the optimal scheduling scheme is to minimize the IMG operation costs $F_1$, which can be formulated as [8]

$$\min F_1 = F_{C\_cd} + F_{C\_sr} + F_{C\_fu} + F_{C\_re} \quad (8)$$

where $F_{C\_cd}$, $F_{C\_sr}$, $F_{C\_fu}$, and $F_{C\_re}$ represent respectively charge-discharge cost, SR cost, the fuel cost of a MT unit, and the cost of reserve provided by the ESS.

$$F_{C\_cd} = \sum_{t=1}^{T}\left(\omega_{rt\_price,t} \times \left(P_t^{DC} - P_t^{CH}\right)\right) \quad (9)$$

$$F_{C\_sr} = \sum_{t=1}^{T}\sum_{n=1}^{M_G}\varsigma_n R_{n,t}^{MT} \quad (10)$$

$$F_{C\_fu} = \sum_{t=1}^{T}\sum_{n=1}^{M_G}\kappa_n S_{n,t} + U_{n,t}\left(\zeta_n + \psi_n P_{n,t}^{MT}\right) \quad (11)$$

$$F_{C\_re} = \omega_{rc\_price} \times \sum_{t=1}^{T}P_{Ress,t} \quad (12)$$

Eq. (9)-(12) are respectively the detailed formulations of the charge-discharge cost, the SR cost provided by MTs, the fuel cost of MTs, and the SR cost provided by ESS, where $T$ and $t$ respectively denote an entire scheduling cycle and a time period (in hours). In Eq. (9),

$\omega_{rt\_price,t}$ is the TOU price during period $t$, $P_t^{CH}$ and $P_t^{DC}$ are the ESS charge-discharge powers. In Eq. (10), $M_G$ denotes the number of MT units, $\varsigma_n$ and $R_{n,t}^{MT}$ respectively represent the SR cost and the spinning reserve capacity of MT $n$. In Eq. (11), $\zeta_n$ and $\psi_n$ are the consumption coefficients of MT $n$, $U_{n,t}$ and $S_{n,t}$ are binary variables which denote the operating status and start-stop of the $n$th MT, $\kappa_n$ is the start-stop cost, $P_{n,t}^{MT}$ represents the MT output power during period $t$. In Eq. (12), $\omega_{rc\_price}$ denotes the SR price of the ESS, $P_{Ress,t}$ is the reserve capacity provided by ESS.

*3.1.2 Gas Emission:* Research shows that the main pollution gases from MTs include $SO_2$, $NO_x$, $CO_2$, and CO [31]. As a mainly considered goal, the gas emission model is formulated by

$$\min F_2 = \sum_{t=1}^{T}\sum_{k=1}^{K}\left(\sum_{n=1}^{M_G} a_{n,k} P_{n,t}^{MT}\right) \quad (13)$$

The objective function includes additional variables $k$ and $a_{n,k}$, where $k$ represents different kinds of pollution gases, $a_{n,k}$ is the emission coefficient of the $k$th pollution gas of the $n$-th microturbine.

*3.1.3 Consumer Satisfaction:* Previous investigations suggest that a good dispatch scheme for commercial microgrids must be addressed towards two principal aims: customer satisfaction and cost efficiency [32, 33].

To evaluate the quality of user experience, from the perspective of demand side management a new consumer satisfaction indicator $F_3$ is developed as follows.

$$\max F_3 = \frac{\sum_{t=1}^{T}\left(\sum_{n=1}^{M_G} P_{n,t}^{MT} + P_t^{PV} + P_t^{WT} + P_t^{DC} - P_t^{CH}\right)}{\sum_{t=1}^{T} P_t^L} \times 100\% \quad (14)$$

Eq. (14) describe the consumer satisfaction through the ratios of total power generation and total loads during all periods, where $P_t^{WT}$ and $P_t^{PV}$ are respectively the output of WT and PV during period $t$, $P_t^L$ is the load power during period $t$.

### 3.2. Constraints

The constraints required to the IMG system operation are shown as follow:

$$\sum_{n=1}^{M_G} P_{n,t}^{MT} + P_t^{DC} - P_t^{CH} = E(P_t^{EL}) + P_t^{CNLOAD}, \forall t \quad (15)$$

$$E(P_t^{EL}) = \sum_{u_{d,t}=0}^{N_{d,t}} u_{d,t} qa(u_{d,t}) - \sum_{u_{a,t}=0}^{N_{a,t}} u_{a,t} qa(u_{a,t}) - \sum_{u_{b,t}=0}^{N_{b,t}} u_{b,t} qa(u_{b,t}) \quad (16)$$

$$U_{n,t} P_{n,min}^{MT} \leq P_{n,t}^{MT} \leq U_{n,t} P_{n,max}^{MT}, \forall t, n \in M_G \quad (17)$$

$$\begin{cases} C_{t+1} = C_t + \eta^{CH} P_t^{CH} \Delta t \\ C_{t+1} = C_t - \Delta t P_t^{DC} / \eta^{DC} \end{cases}, \forall t \quad (18)$$

$$\begin{cases} 0 \leq P_t^{CH} \leq P_{max}^{CH} \\ 0 \leq P_t^{DC} \leq P_{max}^{DC} \end{cases} \forall t \quad (19)$$

$$C_{min} \leq C_t \leq C_{max}, \forall t \quad (20)$$

$$P_{n,t}^{MT} + R_{n,t}^{MT} \leq U_{n,t} P_{n,max}^{MT}, \forall t, n \in M_G \quad (21)$$

$$P_{Ress,t} \leq \min\{\eta_{dc}(C_t - C_{min})/\Delta t, P_{max}^{DC} - P_t^{DC}\}, \forall t \quad (22)$$

$$Prb\left\{\sum_{n=1}^{M_G} R_{n,t}^{MT} + P_{Ress,t} \geq (P_t^L - P_t^{WT} - P_t^{PV}) - E(P_t^{EL})\right\} \geq \alpha, \forall t \quad (23)$$

Eq. (15)-(16) and Eq. (17) are respectively the system power balance constraint and the MT power output constraint, where $P_t^{CH}$ and $P_t^{DC}$ are respectively the ESS charge and discharge powers, and they both are collectively called exchanging powers. $P_t^{CNLOAD}$ is the controllable load output, and $E(P_t^{EL})$ is the expected value of $P_t^{EL}$, $P_{n,min}^{MT}$ and $P_{n,max}^{MT}$ are respectively the minimal and maximum values of the $n$th MT's output.

Eq. (18)-(20) are the constraints related to the ESS. Eq. (18) is the charge-discharge constraint, where $C_{t+1}$ and $C_t$ are the available capacity of the ESS during period $t+1$ and $t$, $\eta^{DC}$ and $\eta^{CH}$ denote the discharge and charge efficiencies of the ESS. Eq. (19) is the charge-discharge rate limits, $P_{max}^{CH}$ and $P_{max}^{DC}$ are the maximum values of the ESS charge and discharge powers. Eq. (20) is the capacity limits, $C_{min}$ and $C_{max}$ are the minimal and maximum energies stored in ESS.

Eq. (21)-(23) describe the constraints regarding spinning reserves. Concretely speaking, Eq. (21) is the constraint of spinning reserves provided by the MT units, while Eq. (22) expresses that of the ESS, where $P_{Ress,t}$ denotes the ESS reserve capacity in period $t$. In general, the operation of IMG is susceptible to uncertainties from both sources and loads since it is unable to obtain power supports from the main grid and relatively small capacity [8]. In terms of the IMG studied in this work, there are multiple DGs with different probability distribution characteristics on the source side, while load fluctuations increase the uncertainty. And thereby, sufficient spinning reserves are required to maintain the system reliable operation [29], which will inevitably lead to an expensive cost. Eq. (23) describes the probabilistic spinning reserve requirement, in which α is the pre-specified confidence threshold.

### 4. Solution methodology

In this section, the proposed two-step approach for solving the built multi-objective scheduling model is described in detail. First, the serialization description of random variables is introduced; and then, treatment of chance constraints chance constraint into its deterministic form is described; next, the basic principles of the θ-DEA, fuzzy C-means clustering, and grey relation projection are briefly presented; what's more, how to calculate the optimal scheme is given, and finally, the specific solving process are listed. It should be noted that the decision-maker refers to the operation dispatcher of the microgrid in this work.

### 4.1. Serialization description of random variables

*4.1.1 Introduction of sequence operation theory (SOT):* SOT proposed by Prof. Kang is an effective tool for addressing the uncertainties of power systems [34]. In this theory, the probability distributions of random variables are denoted by probabilistic sequences (PSs), and a new sequence could be derived from the operations between sequences. And then, the new probability distributions of random variables are obtained through mutual operations. Concretely speaking, four typical kinds of operations have been defined. The details of the theory are given in [34].

*4.1.2 Probability sequences of DG outputs:* In this work, all DG outputs are formulated by the probabilistic sequences by the discretizing continuous probability distributions. More specifically, the WT and PV are respectively depicted by using the probabilistic sequences $a(i_a)$ with the length $N_a$ and $b(i_b)$ with the length $N_b$.

The probabilistic sequences of the DG outputs can be respectively obtained as follows:

$$a(i_{a,t}) = \begin{cases} \int_0^{q/2} f_o(P^{WT})dP^{WT}, & i_{a,t}=0 \\ \int_{i_{a,t}q-q/2}^{i_{a,t}q+q/2} f_o(P^{WT})dP^{WT}, & i_{a,t}>0, i_{a,t}\neq N_{a,t} \\ \int_{i_{a,t}q-q/2}^{i_{a,t}q} f_o(P^{WT})dP^{WT}, & i_{a,t}=N_{a,t} \end{cases} \quad (24)$$

$$b(i_{b,t}) = \begin{cases} \int_0^{q/2} f_p(P^{PV})dP^{PV}, & i_{b,t}=0 \\ \int_{i_{a,t}q-q/2}^{i_{a,t}q+q/2} f_p(P^{PV})dP^{PV}, & i_{b,t}>0, i_{b,t}\neq N_{b,t} \\ \int_{i_{a,t}q-q/2}^{i_{a,t}q} f_p(P^{PV})dP^{PV}, & i_{b,t}=N_{b,t} \end{cases} \quad (25)$$

where $q$ is the pre-given discretization step.

### 4.2. Treatment of chance constraints
*4.2.1 Probabilistic sequence of EL power:* According to the SOT, the probabilistic sequence $c(i_{c,t})$ of the joint power outputs of WT and PV units is

$$c(i_{c,t}) = \sum_{i_{a,t}+i_{b,t}=i_{c,t}} a(i_{a,t})b(i_{b,t}), \ i_{c,t}=0,1,...,N_{a,t}+N_{b,t} \quad (26)$$

Given the PS of the load power $d(i_{d,t})$, the PS of the EL power $e(i_{e,t})$ can be obtained as follows [8]:

$$e(i_{e,t}) = \begin{cases} \sum_{i_{d,t}-i_{c,t}=i_{e,t}} d(i_{d,t})c(i_{c,t}), & 1 \leq i_{e,t} \leq N_{e,t} \\ \sum_{i_{d,t} \leq i_{c,t}} d(i_{d,t})c(i_{c,t}), & i_{e,t}=0 \end{cases} \quad (27)$$

The PS of the EL power is listed in the following table.

**Table 1** PS of the EL power

| Power (kW) | 0 | $q$ | … | $u_e q$ | … | $N_{e,t} q$ |
|---|---|---|---|---|---|---|
| Probability | $e(0)$ | $e(1)$ | … | $e(u_e)$ | … | $e(N_{e,t})$ |

Table 1 suggests that there exists a one-to-one correspondence between the EL power $u_e q$ and the corresponding probability $e(u_e)$. Similarly, all such probabilities during different time periods can form a probabilistic sequence $e(i_{e,t})$.

*4.2.2 Treatment of chance constraints:* For purpose of handling the chance constraint in Eq. (23), a 0-1 variable $W_{u_{e,t}}$ is introduced as follows [8].

$$W_{u_{e,t}} = \begin{cases} 1, & \sum_{n=1}^{M_G} R_{n,t}^{MT} + P_{Ress,t} \geq u_{e,t}q - E(P_t^{EL}) \\ 0, & otherwise \end{cases} \forall t, u_{e,t}=0,1,...,N_{e,t} \quad (28)$$

And then, Eq. (23) can be rewritten as

$$\sum_{u_{e,t}=0}^{N_{e,t}} W_{u_{e,t}} e(u_{e,t}) \geq \alpha \quad (29)$$

It can be observed From Eq. (29) that for all the possible EL outputs, the spinning reserve capacities in the MG satisfy the following condition: the required confidence is not less than the confidence threshold α. Consequently, it can be derived that (29) is equivalent to (23). In this way, the tricky the chance constraint has been transformed into its equivalent deterministic form.

### 4.3. θ-DEA algorithm
The $θ$-DEA, initially put forward by Yuan in 2016 [35], is a novel powerful approach to address a MOO issue, which has been successfully employed to solve many engineering problems [31].

*4.3.1 Basic principles:* The $θ$-DEA belongs to a multiobjective evolutionary algorithm (MOEA) on the basis of a new dominance relation, called $θ$-dominance. In $θ$-DEA, the concept of $θ$-dominance is of the utmost importance to handle the balance between convergence and diversity. Due to space limitations, a brief description of the $θ$-dominance is given as follow.

Assuming every solution belonging to population $S$ is related to a cluster through clustering operators, $\tilde{F}(x)=(f_1(x), f_2(x), ..., f_M(x))^T$ denotes the normalized objective vector associated with solution $x$, $L$ refers to the line which passes through the two points: $\lambda_j$ (the $j$th reference point) and the origin, $\mathbf{u}$ is the projection of $\tilde{F}(x)$ on $L$. Let

$$\Upsilon_j(x) = Dis_{j,1}(x) + \theta \times Dis_{j,2}(x), \ j \in \{1,2,...,N_{cl}\} \quad (30)$$

where $θ$ and $N_{cl}$ are the penalty parameter and the cluster number counts, $Dis_{j,1}(x)$ and $Dis_{j,2}(x)$ denote the distances between $\mathbf{u}$ and the origin, $\tilde{F}(x)$ and $L$, respectively.

Taking 2-D objective space as an example, the above two distances can be shown in Fig. 1.

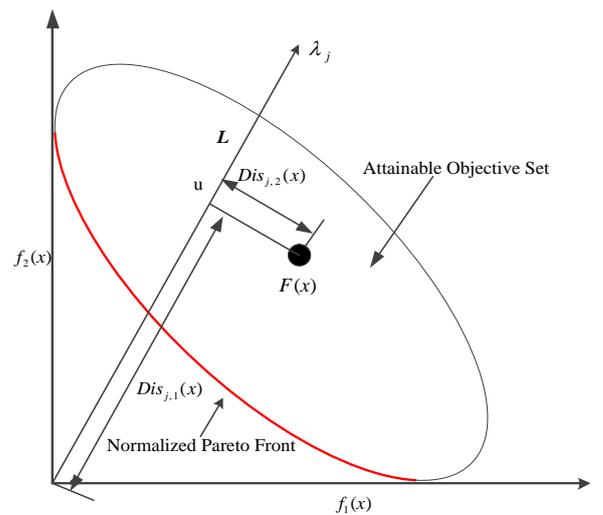

**Fig. 1.** *Schematic of the distances in 2-D objective space*

**Definition [35]:** Assuming two solutions $x_1$, $x_2$ belonging to a cluster, $x_1$ is said to $θ$-dominate $x_2$, denoted by $x_1 \prec_\theta x_2$, and $\Upsilon_j(x_1) < \Upsilon_j(x_2)$, where $j \in \{1,2,\cdots,N_{cl}\}$.

Investigations suggest that the $θ$-DEA is capable of handling the trade-off between the algorithm's convergence and diversity. More details regarding this algorithm can be found in the related references [31, 35].

*4.3.2 Hybrid coding:* Considering the characteristics of controlled variables, a hybrid real/integer-coded strategy is adopted in this study [36]. The continuous variables comprise the $n$th ($n \in M_G$) MT's active power outputs $P_n^{MT}$, the reserve capacity $R_n^{MT}$, the reserve capacity provided by the ESS $P_{Ress}$, while the discrete variables comprise the state variables $U_n$ and start-up variables $S_n$.

*4.4. Fuzzy C-means Clustering*

In this work, FCM clustering is adopted to divide the POSs into different clusters. It is a famous unsupervised technique that seeks to solve the following optimization problem [36]:

$$\min J(W,U,V) = \sum_{i=1}^{N_{po}} \sum_{j=1}^{N_{nu}} \mu_{ij}^m \|w_i - v_j\|^2$$
$$s.t. \quad \sum_{j=1}^{N_{cl}} \mu_{ij} = 1 \quad (31)$$

where $J$ is a loss function, $N_{po}$ and $N_{nu}$ are the number of the POSs and clusters; $W$, $U$, and $V$ respectively denote an input vector, the membership degree matrix, and its cluster centers; $\mu_{ij}$ is the degree of membership; $v_j$ denotes the $i$th cluster, $m$ is a control parameter regarding the fuzziness degree.

*4.5. Grey Relation Projection*

The GRP theory is a powerful technique to analyze the relationship between decisions with grey information, and it has been widely utilized in many engineering areas [31, 36]. The projection $Pro_h$ of a scheduling scheme on an ideal scheme is expressed as

$$Pro_h^{+(-)} = \sum_{y=1}^{N_g} Grc_{h,y}^{+(-)} \frac{w_y^2}{\sqrt{\sum_{k=1}^{N_g}(w_y)^2}} \quad (32)$$

where "+"/ "−" respectively denote a positive/negative scheme, $Grc_{h,y}$ denotes the grey relation coefficient between scheme $h$ and indicator $y$, $N_g$ is the total number of indicators, $w_y$ is the weight coefficient of indicator $y$.

And then, the BCSs are identified according to the following relative projection value (RPV) [36]:

$$RPV_h = \frac{Pro_h^+}{Pro_h^+ + Pro_h^-}, \quad 0 \le RP_h \le 1 \quad (33)$$

where $RPV_h$ denotes the RPV of scheme $h$. It should be noted that the solutions with the highest RPVs will be chosen as the BCSs.

*4.6. Solving process*

The solving process of the proposed approach can be divided into the follows steps.
Step 1: Model the IMG according to (8) ~ (23).
Step 2: Handle the chance constraints.
Step 3: Obtain the scheme model with a MILP form.
Step 4: Input the parameters of the IMG and TOU price.

Step 5: Search the POSs of the problem by using the *θ*-DEA.
Step 6: Check the existence of a solution. If a solution is found, continue the optimization process; otherwise, update the relevant parameter, and thereby go to Step 5.
Step 7: Obtain the Pareto optimal solutions.
Step 8: Initialize the matrix $U$.
Step 9: Compute the cluster centers $V$.
Step 10: Check whether the inequality $|J_{cur} - J_{pre}| < \varepsilon$ is satisfied, where $J_{cur}$ and $J_{pre}$ denote the values of function $J$ at the current and previous iteration. If satisfied, update $U$ and return to Step 8; otherwise, output the clusters.
Step 11: Build an initial decision matrix.
Step 12: Standardize the decision matrix.
Step 13: Calculate the coefficient $Grc$.
Step 14: Calculate the projection $Pro_h$.
Step 15: Calculate the $RPV_h$.
Step 16: Output the BCSs.
Step 17: Obtain the optimal scheduling schemes of IMG.

For ease of description, Fig. 2 shows the above-mentioned solving process.

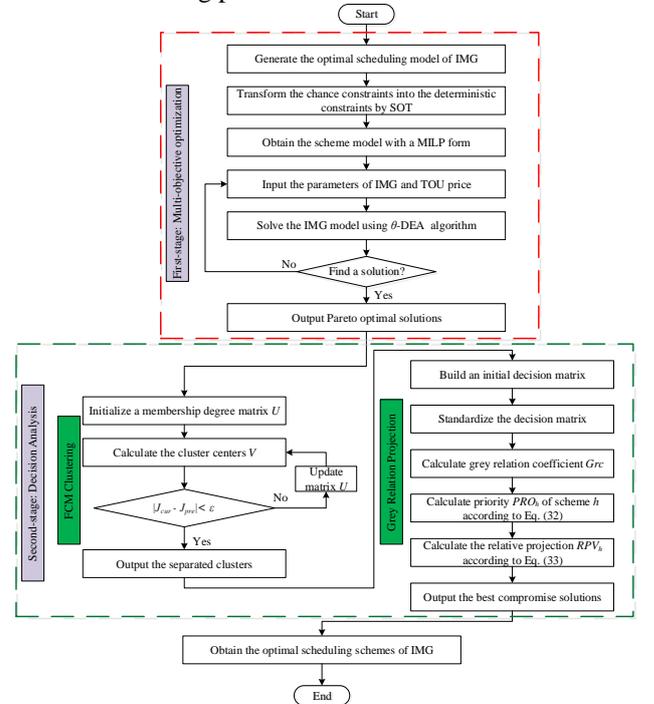

**Fig. 2.** *Solution framework of the proposed approach*

## 5. Case Study

The proposed method is examined on a modified microgrid test system which was originally proposed by the Distributed Energy Control and Communication (DECC) lab at Oak Ridge National Laboratory (ORNL) [8, 37]. All simulation tests are carried out under the MATLAB environment on a desktop computer that is configured with Intel Core dual-core processors at 2.4 GHz and 6 GB RAM.

*5.1. Testing system*

Fig. 3 shows the structure of the used testing system, which is composed of a photovoltaic panel, a wind turbine, three MT units, and an ESS, where PCC denotes the point of common coupling.

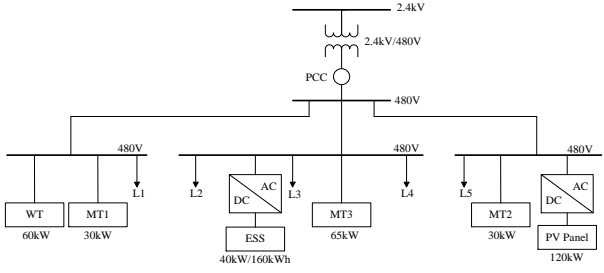

**Fig. 3.** *Microgrid testing system*

The relevant parameters of the MG model, such as MT units, WT, PV, and ESS, are detailedly given in [8]. In addition, the price of reserve capacity provided by the ESS is set as $\omega_{rc\_price}$ =0.02 $/kW.

The gas emission coefficients of the MTs used in this study are shown in Table 2.

**Table 2** Emission coefficients of the MT units

| Gas classification | Emission coefficients (g/kWh) | | |
|---|---|---|---|
| | MT1 | MT2 | MT3 |
| $NO_X$ | 0.619 | 4.33 | 0.023 |
| $CO_2$ | 184 | 232 | 635 |
| CO | 0.17 | 2.32 | 0.054 |
| $SO_2$ | 0.000928 | 0.00464 | 0.0012 |

In this study, the used TOU prices between IMG and ESS are listed in Table 3.

**Table 3** TOU electrical prices

| Time periods | Specific times | Price($/kWh) | |
|---|---|---|---|
| | | Purchase | Sale |
| Peak period | 11:00-15:00 | 0.83 | 0.65 |
| flat period | 00:08-11:00,15:00-19:00,20:00-24:00 | 0.49 | 0.38 |
| Off-peak period | 00:00-00:08, 19:00-20:00 | 0.17 | 0.13 |

### 5.2. Analysis and Discussion

For ease of analysis without loss of generality, these studies were conducted at a discrete step of 2.5 kW, a confidence level of 90%, and a load fluctuation of 10 %. The population size of the $\theta$-DEA algorithm is set to 100, and the maximum iteration number is 100.

*5.2.1 Optimization Results:* After multiple iterations, the Pareto-optimal solutions are obtained by the $\theta$-DEA, which is shown in Fig. 4.

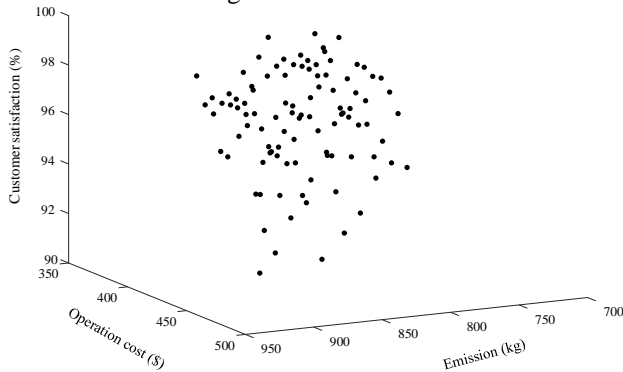

**Fig. 4.** *Distribution of Pareto-optimal solutions*

Fig. 4 indicates that it is distinct that the method proposed can obtain almost complete and uniform Pareto optimals. Thus, it can be concluded that the method proposed can coordinate multi-objectives in IMG. The extreme solutions from Pareto-optimal solutions are shown in Table 4.

**Table 4** Extreme solutions of the obtained Pareto-optimal solution set

| Extreme solutions | $F_1$($) | $F_2$(kg) | $F_3$(%) |
|---|---|---|---|
| Extreme solution 1 | 370.11 | 765.97 | 90.5 |
| Extreme solution 2 | 379.19 | 727.96 | 93.1 |
| Extreme solution 3 | 452.85 | 899.54 | 100 |

As shown in Table 4, the extreme solutions 1 and 2 respectively reach the minimum values of the objective function $F_1$ and $F_2$, and the extreme solution 3 reach the maximal value of objective function $F_3$. Note that, the obtained every extreme solutions are optimal for a single-objective optimization; but for multi-objective optimization, they are non-inferior solutions.

The POSs obtained above are clustered into three groups via FCM clustering algorithm and named for different types with different colors. The clustering results are shown in Fig. 5. It should be noted that in this work, types A, B, and C respectively reflect decision-makers' preferences on the objective functions $F_3$, $F_1$ and $F_2$.

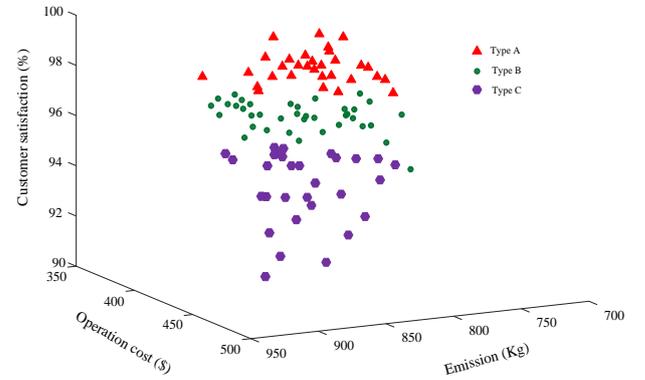

**Fig. 5.** *Distribution of Pareto-optimal solutions after clustering*

After clustering, the priority memberships are calculated by GRP in the same groups, and the solutions with the highest priority memberships are considered as the BCSs, as shown in Table 5.

**Table 5** Best compromise solutions

| Compromise solutions | $F_1$($) | $F_2$(kg) | $F_3$(%) | Priority membership |
|---|---|---|---|---|
| BCS 1 | 389.61 | 813.62 | 95.9 | 0.7314 |
| BCS 2 | 406.05 | 794.68 | 97.6 | 0.6963 |
| BCS 3 | 448.95 | 890.01 | 99.1 | 0.7589 |

As shown in Table 5, different BCSs can be chosen according to decision makers' preferences. Concretely speaking, if the economy index $F_1$ is placed as the highest priority, then the BCS 1 will be considered as the best one; if the environmental protection is chosen as the primary concern, the BCS2 is no doubt the optimal choice, since which seeks to the minimization of pollutant emissions; if a decision maker puts more emphasis on user experience, the BCS3 will be the best choice. By this means, the BCSs can be automatically identified by using the FCM-GRP decision analysis, which is helpful for providing more realistic options to a decision maker.

*5.2.2 Dispatch schemes corresponding to different BCSs:* Assuming the load standard deviation 10% and the confidence level 90%, the obtained optimal dispatch schemes corresponding to the BCSs are shown in Figs. 6-8.

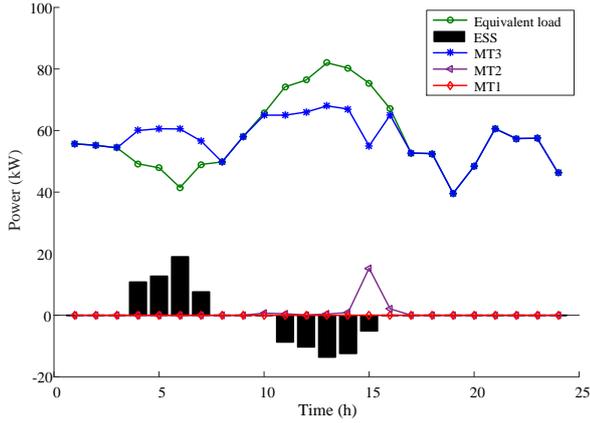

**Fig. 6.** *Optimal dispatch scheme corresponding to BCS 1*

From Fig. 6, it can be observed that MT3 is given as a priority to output power to isolated microgrid in BCS 1. The reason is that the decision-makers are prone to emphasize the economy, while the operation cost of MT3 is lower than the other MTs in BCS 1.

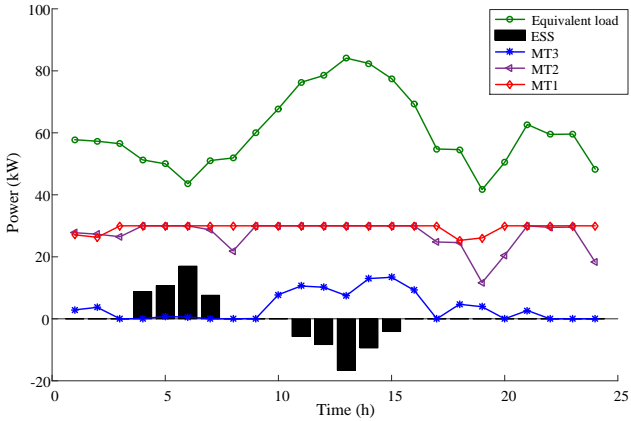

**Fig. 7.** *Optimal dispatch scheme corresponding to BCS 2*

Fig. 7 shows that in BCS2, MT1 and MT2 are used as main units that supply powers to the isolated microgrid. The reason is that the environmental factor is considered as a primary objective in this case, while the gas emission factors of MT1 and MT2 are less than that of MT3. Under the premise of ensuring the balance between supply and demand, the increase of the outputs of MT2 and MT3 effectively reduces the total pollution gas emissions.

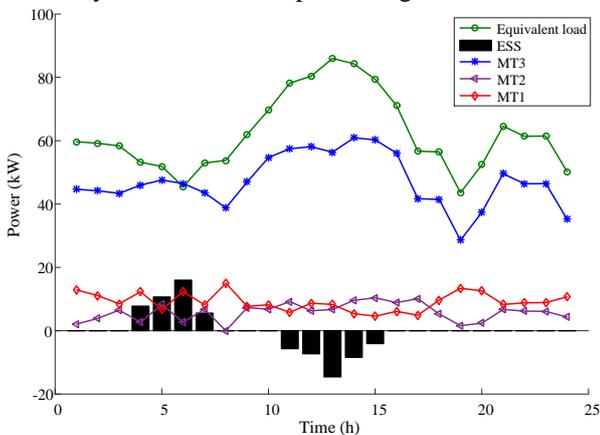

**Fig. 8.** *Optimal dispatch scheme corresponding to BCS 3*

Fig. 8 shows the output of three MTs and ESS during all periods in BCS 3. In this case, the consumer satisfaction is chosen as the principal objective, and the total outputs of MTs and ESS are the highest than those in the other two cases, since, according to Eq. (14), the total power supply from MTs and energy storage should be as great as possible to maximize the consumer satisfaction indicator $F_3$. Therefore, from Figs. 6-8, it can be seen that difference in the outputs of MTs mainly depend on the decision maker's preferences in different BCSs.

*5.2.3 Reserve capacity analysis:* Reserve capacity is an important mean to maintain the balance between supply and demand of the system [29].

Fig. 9 gives the required total spinning reserves under several confidence levels.

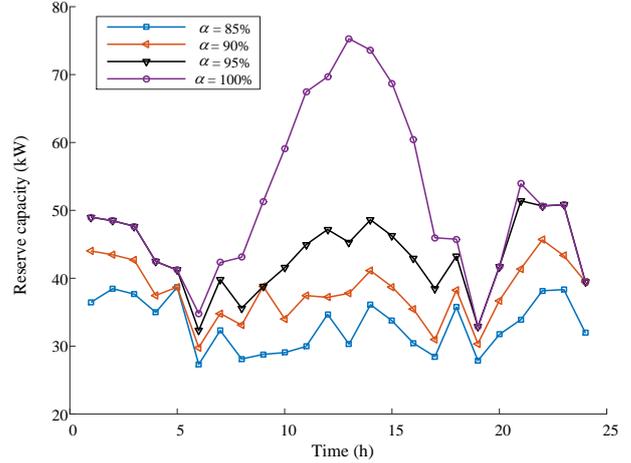

**Fig. 9.** *Total spinning reserves under different confidence levels*

It can be observed in Fig. 9 that, with the increase of the confidence levels, the reserve capacity required for the system is gradually increasing, which inevitably increases the operating costs. The main reason is that the greater the confidence level, the more reserve the system needs to counteract the fluctuation of renewable energy output. Therefore, it is of critical importance to select the appropriate confidence level to achieve a better balance between reliability and economy.

To further analyze the spinning reserves provided from different sources, taking the confidence level $α=90\%$ as an example, the total spinning reserves and the reserves provided by the MTs and ESS in different BCSs are illustrated in the following figure.

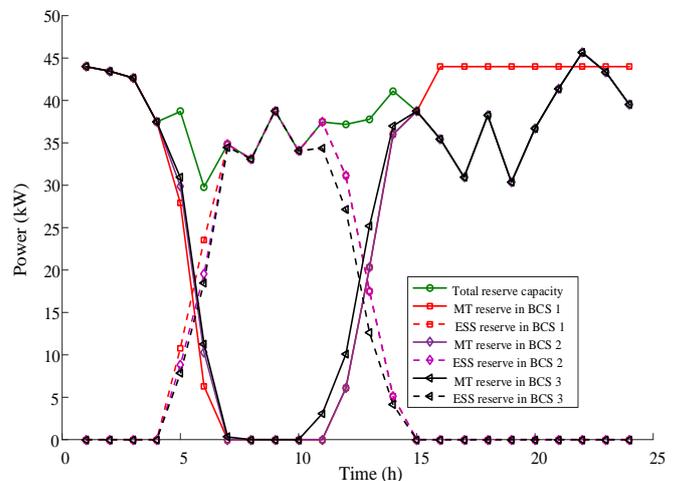

**Fig. 10.** *Reserve capacities provided by the ESS and MTs in different BCSs*

It can be seen from Fig. 10 that in most time periods the reserve capacity from the ESS is greater than that from the MTs. The reason for this is that the ESS is a more preferable provider due to its lower cost and faster response [38, 39]. Only in the case that the ESS dump energy is unable to meet the required reserves alone, the MTs are used to provide this service. Therefore, besides dispatchable generators like MT units, the ESS is capable of supplying spinning reserve services for isolated MGs.

*5.2.4 Computational efficiency analysis:* In order to properly evaluate the computational efficiency of the proposed approach, the used calculation time is demonstrated in Table 6.

**Table 6** Calculation time of the proposed approach

| Items | Time (s) |
|---|---|
| Multi-objective optimization (1st step) | 294.2 |
| Decision analysis (2nd step) | 1.6 |
| Total time | 295.8 |

From Table 6, it can be observed that the total computation times of the proposed method are less than 300 seconds. It can be expected that if more advanced hardware configurations are used, the computational efficiency will be further improved. Therefore, the conclusion can be safely drawn that the computational efficiency of the proposed method meets the real-time requirements of the microgrid scheduling.

## 6. Conclusion

The purpose of this paper is to present a multi-objective dynamic optimal dispatch model for coordinating the economy, environmental protection and user experience of isolated microgrids. To solve this model, a two-step solution approach is proposed by integrating multi-objective optimization and decision analysis. Simulation results suggest that the approach not only can yield multiple well-distributed Pareto optimal solutions, but also can identify the BCSs representing decision makers' different preferences automatically. In addition, the computational efficiency of our method satisfies the real-time requirements of microgrid scheduling.

Future work will focus on applying demand responses in the field of microgrid dispatch to address the renewables variability and alleviate the grid pressure during on-peak periods [13, 40, 41]. Besides, it is interesting to investigate the optimal dispatch of islanded microgrids under unbalanced three-phase conditions [42, 43]. It is another potential future research topic concerning applying big data and machine learning techniques to develop a model-free optimal demand response scheduling methodology for Microgrids [44].

## 7. Acknowledgements

This work is supported by the China Scholarship Council (CSC) under Grant No. 201608220144.